\documentclass[12pt]{iopart}

\usepackage{graphicx}
\usepackage{dcolumn}
\usepackage{bm}
\usepackage{hyperref}
\usepackage[dvipsnames]{xcolor}

\makeatletter
\@namedef{ver@amsmath.sty}{}
\makeatother
\usepackage{amstext}
\usepackage{chemformula}
\usepackage[T1]{fontenc}       

\begin{document}

\title{Magnetic Skyrmion Annihilation by Quantum Mechanical Tunneling}

\author{Sergei M. Vlasov$^{a,b}$, Pavel F. Bessarab$^{a,b,c}$, Igor S. Lobanov$^{b}$, Mariia N. Potkina$^{d}$, Valery M. Uzdin$^{b,d}$ and Hannes J\'onsson$^{a,e}$}

\address{$^1$Science Institute and Faculty of Physical Sciences, University of Iceland VR-III, 107 Reykjav\'{\i}k, Iceland}
\address{$^2$ITMO University, 197101 Saint Petersburg, Russia }
\address{$^3$Peter Gr\"unberg Institute and Institute for Advanced Simulation, Forschungszentrum J\"ulich, 52425 J\"ulich, Germany}
\address{$^4$Department of Physics, St. Petersburg State Univ., St. Petersburg, 198504 Russia}
\address{$^5$Department of Applied Physics, Aalto University, Espoo, FI-00076, Finland}

\eads{\mailto{hj@hi.is}}

\begin{abstract}
Magnetic skyrmions are nano-scale magnetic states that could be used in various spintronics devices. A central issue is the mechanism and rate of various possible annihilation processes and the lifetime of metastable skyrmions. While most studies have focused on classical over-the-barrier mechanism for annihilation, it is also possible that quantum mechanical tunneling through the energy barrier takes place. Calculations of the lifetime of magnetic skyrmions in a two-dimensional lattice are presented and the rate of tunneling compared with the classical annihilation rate. A remarkably strong variation in the crossover temperature and the lifetime of the skyrmion is found as a function of the values of parameters in the extended Heisenberg Hamiltonian, i.e. the out-of-plane anisotropy, Dzyaloshinskii-Moriya interaction (DMI) and applied magnetic field. Materials parameters and conditions are identified where the onset of tunneling could be observed on a laboratory time scale. In particular, it is predicted that skyrmion tunneling could be observed in the PdFe/Ir(111) system when an external magnetic field on the order of 6 $T$ is applied.
\end{abstract}

\maketitle





\section{Introduction}

Non-collinear localized magnetic states are receiving a great deal of attention, in particular magnetic skyrmions which have been proposed as elements in
future spintronics devices~\cite{Kiselev_2011,Fert_2013,Fert_2017}.
Along with interesting transport properties, skyrmions exhibit particle-like behavior and carry a topological charge enhancing their stability with respect to a uniform  ground state, typically a ferromagnetic phase.
A key issue is the lifetime of the skyrmions and its dependence on temperature and applied magnetic field.
Various mechanisms for the annihilation of a skyrmion have been characterized by theoretical modeling of atomic scale systems, in particular
collapse in the interior of the sample~\cite{Bessarab_2015,Lobanov_2016,Stosic_2017,Uzdin_2018} 
and escape through the boundary of the magnetic domain~\cite{Stosic_2017,Uzdin_2018,Bessarab_2018}. 
Two skyrmions can also merge into one, as well as the reverse process where a skyrmion is transformed into a pair of identical skyrmions~\cite{Muller_2018}.

The activation energy for the various possible transitions can be calculated for a given spin Hamiltonian by finding the minimum energy path (MEP) 
from the local energy minimum corresponding to the skyrmion state to the final state minimum corresponding to the uniform state.
The maximum in energy along the path corresponds to a first order saddle point on the energy surface.
The MEP can be found using the geodesic nudged elastic band method~\cite{Bessarab_2015} 
and the calculation accelerated by making use of knowledge obtained 
from previous calculations by focusing only on the region near the maximum~\cite{Lobanov_2017}. 
For the classical over-the-barrier mechanism, the pre-exponential in the Arrhenius type rate expression can be estimated using 
harmonic transition state theory (HTST) for magnetic systems~\cite{Bessarab_2012,Bessarab_2013}.
This has, for example, been done for skyrmions 
in a PdFe overlayer on Ir(111) surface~\cite{Bessarab_2018,Uzdin_2018}, 
a system that has been studied extensively in the laboratory~\cite{Romming_2013,Hagemeister_2015}.
The challenge is to design materials where skyrmions are sufficiently stable at room temperature and still small enough to be used in 
spintronic devices.
Theoretical calculations can help accelerate this development by 
identifying the dominant annihilation mechanisms and
predicting how the stability of skyrmions depends on materials properties.
%

In most calculations it is assumed that the system is able to overcome the energy barrier of the transition by thermal activation and this is typically a valid
assumption. 
It is, however, possible that quantum mechanical tunneling brings the system from the metastable skyrmion state to the ground state. 
Tunneling in systems described by a single magnetic moment, within a macrospin approximation, has been studied extensively 
\cite{Gunther_1995,Chudnovsky_1998,Garanin_1999,Garanin_2000,Vlasov_2016,Vlasov_2017a,Vlasov_2017b}, 
in particular in the context of molecular magnets, both experimentally~\cite{Gatteschi_2006,Gatteschi_2003,Aubin_1998,Wernsdorfer_2002} 
and theoretically~\cite{Chudnovsky_1998,Vlasov_2017a}.
For example, the rate of magnetization reversals in a Mn$_4$ monomer and dimer  
molecular magnets has been calculated as a function of temperature and excellent agreement obtained
with the experimentally measured rates~\cite{Aubin_1998,Wernsdorfer_2002} 
both the high temperature classical regime and the crossover temperature for tunneling, 
using a Hamiltonian parametrized from various experimental observations~\cite{Vlasov_2016,Vlasov_2017a}.
 
Since skyrmion stability is of central importance, it is important to have a way to estimate whether tunneling is an important annihilation mechanism.
An experimental observation of skyrmion tunneling would, furthermore, be an example of what is referred to as macroscopic quantum tunneling 
and is of considerable interest in the study of quantum phenomena.

Recently, the quantum mechanical nature of skyrmions has received some attention.
Rold\'an-Molina {\it et al.}~\cite{Roldan_2015} calculated the zero-point energy associated with quantum spin fluctuations 
and found that it can contribute up to 10\% of the 
total Zeeman energy necessary to remove a skyrmion with an applied magnetic field.
Diaz and Arovas~\cite{Diaz16} considered a model where skyrmions are formed by adding a local magnetic field over a circular spot, opposing a uniform magnetic 
field stabilizing the ferromagnetic phase so as to form a single skyrmion ground state. 
The rate of skyrmion nucleation by tunneling at zero temperature to this induced ground state was calculated using path integrals 
and a collective coordinate approximation.
Psaroudaki {\it et al.}~\cite{Psaroudaki_2017} generalized the micromagnetic equations of motion to finite temperature using a path integral formalism and predicted
a quantum mechanical addition to the effective mass of the skyrmion.
Derras-Chouk, Chudnovsky and Garanin~\cite{Derras-Chouk_2018} derived a Lagrangian describing the coupled dynamics of skyrmion radius and chirality angle 
and estimated the tunneling rate of skyrmion collapse within an instanton approach.
In their calculations the skyrmion is described as a Belavin-Polyakov (BP) soliton~\cite{Belavin_1975}, an approximation corresponding
to small DMI and crystalline anisotropy as compared to the exchange interaction.
The BP soliton shape can be a good approximation for some systems but is not of general validity and the 
widely studied PdFe/Ir(111) system is one example where it does not apply, as illustrated below. 
%
In order to search more generally over materials parameters and experimental conditions to evaluate the possibility of observing skyrmion tunneling
in the laboratory, it is important to use as general approach as possible to predict both the crossover temperature for tunneling 
as well as the lifetime of the skyrmion at that temperature.

This article presents results of calculations of the crossover temperature for quantum mechanical tunneling for a range of materials parameters, with 
special focus on the PdFe/Ir(111) system.
The magnetic moments in the system are taken into account explicitly without introducing collective coordinates {\it a priori}. 
The method is based on instanton theory and represents an extension of methodology presented earlier for systems described 
by a single magnetic moment \cite{Vlasov_2016,Vlasov_2017a}.
A region in parameter space is identified where the lifetime of the skyrmion is on a laboratory time scale at the crossover temperature, thus identifying possible 
candidate materials for the observation of skyrmion tunneling. 
The prediction is made that skyrmion tunneling could be observed 
in PdFe/Ir(111) in the presence of an external magnetic field.


\section{Methods}

Within HTST for over-the-barrier transitions in magnetic systems, 
the mechanism and rate of thermally induced transitions is characterized by the first order saddle point on the energy surface, $\vec{\mathbf{s}}^\dagger$
with energy $E^\dagger$, 
representing the highest point along the MEP connecting the initial state and the final state~\cite{Bessarab_2012,Bessarab_2013}.
At a first order saddle point, the Hessian matrix has one negative eigenvalue.
HTST can be used to estimate the lifetime of a metastable skyrmion state when the dominant annihilation mechanism is over-the-barrier transitions 
as is the case at high enough temperature.
As the temperature is lowered, 
the rate of such transitions drops and eventually the temperature independent
quantum mechanical tunneling becomes the dominant transition mechanism.
The challenge is to estimate the crossover temperature, T$_c$, where tunneling becomes dominant.

Instanton theory can be used to define and evaluate T$_c$~\cite{Miller_1975,Coleman_1977,Benderskii_1994}. 
It corresponds to the highest temperature where a 
periodic solution of the equation of motion in imaginary time exists within an infinitesimal vicinity of the first-order saddle point on the energy surface. 
Such a path is referred to as an instanton and from its period, $\beta$, a corresponding temperature can be found as  T = $\hbar/k_B \beta$.
The appearance of such a path as temperature is lowered can be deduced from an analysis of the 
Euclidean (imaginary-time) action. 
For a system with N spins 
with magnitude $\mu$ and orientation defined by a set of unit vectors $\mathbf{s}_i$, $i=1, \dots , N$,
the action is~\cite{Fradkin_1988}
\begin{equation}
\label{eq: action}
	Q[\vec{\mathbf{s}}, \partial_{\tau}\vec{\mathbf{s}}, \tau] = i\mu\sum_{k=1}^{N}\int_{-\beta/2}^{\beta/2}d\tau\mathbf{A}_k\partial_\tau\mathbf{s}_k +
	\int_{-\beta/2}^{\beta/2}d\tau \mathcal{H}(\vec{\mathbf{s}}) ,
\end{equation}
where $\vec{\mathbf{s}} = \{\mathbf{s}_1, \mathbf{s}_2,\ldots\, \mathbf{s}_N\}$, $\mathbf{s}_k(\tau)$ is a closed trajectory, 
$\mathbf{s}_k(-\beta/2) = \mathbf{s}_k(\beta/2)$, 
and $ \mathcal{H}(\vec{\mathbf{s}})$ is the Hamiltonian. The first term in this equation is the Berry phase and $\mathbf{A}_k$ is referred to as Berry connection~\cite{Auerbach_2012}.
It is related to the area of 
on a sphere for each spin bounded by the trajectory and can be expressed as~\cite{Fradkin_1988}
\begin{equation}
\label{eq: solidangle}
       \mathbf{A}_k\partial_\tau\mathbf{s} =
	2\arctan\left(\frac{\partial_\tau \mathbf{s}\cdot(\mathbf{s}\times\mathbf{s}^0)}{2 + 2\mathbf{s}\cdot\mathbf{s}^0 + \partial_\tau\mathbf{s}\cdot\mathbf{s}^0}\right)
\end{equation}
where
$\mathbf{s}^0$ is 
some reference direction.
%
%
\ 

To find an instanton in the vicinity of $\vec{\mathbf{s}}^\dagger$, the action in Eq.~(\ref{eq: action}) is expanded up to second order
\begin{equation}
	\label{eq: expansion of action}
	Q[\vec{\mathbf{s}}] =  Q^\dagger + \delta Q + \frac{1}{2}\delta^2 Q +\ldots,
\end{equation}
where $\vec{\mathbf{s}} = \vec{\mathbf{s}}^\dagger + \delta\vec{\mathbf{s}}$
 and $Q^\dagger$ = $\beta E^\dagger$.
In addition to the boundary conditions, 
$ \delta \mathbf{s}_k(-\beta/2) = \delta \mathbf{s}_k(\beta/2)$ = 0, 
 a normalization constraint needs to be added for each spin,
$\delta \mathbf{s}_k \cdot \mathbf{s}_k = 0$.
At the saddle point $\delta Q = 0$.
The second order variation of the action in the vicinity of the saddle point can be written as
$$
	\frac{1}{2}\delta^2 Q [\vec{\mathbf{s}}] = i\mu\sum_{k=1}^{N}\int_{-\beta/2}^{\beta/2}d\tau
		\left[\frac{\partial_\tau \delta \mathbf{s}_k\times \mathbf{s}^0}{1 + \mathbf{s}^0\cdot \mathbf{s}^\dagger_k} +  
		\frac{\partial_\tau\delta \mathbf{s}_k\times \mathbf{s}^0}{(1 + \mathbf{s}^0\cdot \mathbf{s}^\dagger_k)^2} 
		+ \frac{\partial_\tau\delta \mathbf{s}_k\times \mathbf{s}^\dagger_k}{(1 + \mathbf{s}^0\cdot \mathbf{s}^\dagger_k)^2}\right]\delta \mathbf{s}_k 
$$
\begin{equation}
	\label{eq: second variation}
                 + \int_{-\beta/2}^{\beta/2} d\tau \left(\delta  \vec{\mathbf{s}} ~ \tilde{\nabla}^2 \mathcal{H} ~ \delta   \vec{\mathbf{s}}\right)
\end{equation}
where
 $\tilde{\nabla}^2 \mathcal{H}$ is the Hessian matrix 
evaluated at $\vec{\mathbf{s}}^\dagger$ and the tilde specifies 
a restriction
to the tangent space~\cite{Muller_2018}. 
The linearized Landau-Lifshitz equation 
in imaginary time
can be obtained by 
setting $\delta^2 Q = 0$ and taking the cross product with $\vec{\mathbf{s}}^\dagger$ from the left 
\begin{equation}
	\label{eq: linearized LL equation}
	\partial_\tau\delta \mathbf{s}_k = \frac{i}{\mu}\,
	 (\delta \mathbf{s}_k \times \tilde{\nabla}^2_k \mathcal{H} \, \delta \mathbf{s})
	,\quad \forall k \in [1..N].
\end{equation}
This can be written in matrix form as
\begin{equation}
	\label{eq: matrix LL equation}
	\frac{\partial}{\partial\tau}\delta \vec{\mathbf{s}} = \frac{i}{\mu}\mathbf{M} ~ \delta \vec{\mathbf{s}}.
\end{equation}
The matrix $\mathbf{M}$ has N pairs of complex conjugate eigenvalues, $\lambda_k$.
One of the pairs, chosen here to correspond to $k$=1, is real valued but the other 
$N - 1$ pairs are purely imaginary $\lambda_k = \pm\mathbf{i}\eta_k$ 
A discussion of the properties of the matrix $\mathbf{M}$ can be found in Ref.~\cite{Rozsa_2018, Schutte_2014}.

The general solution of Eq.~(\ref{eq: matrix LL equation}) has the following form
\begin{equation}
	\label{eq: general solution}
	\delta\vec{\mathbf{s}}(\tau) = \sum_{k=1}^N c_k \mathbf{u}_k e^{\omega_k\tau} + 
	c.c.
\end{equation}
where the $\mathbf{u}_k$ are the eigenvectors and $\lambda_k$ the eigenvalues of $\mathbf{M}$, and 
$c_k$ are the expansion coefficients. 
The real pair of eigenvalues, $\lambda_1$, corresponds to the periodic solution with eigenfrequency $\omega_1 = i\lambda_1/\mu$ and this is the instanton. 
The period of motion along this trajectory relates to the frequency as $1/\beta = \vert\omega_1\vert/2\pi$, and gives the onset temperature for tunneling as
\begin{equation}
	\label{eq: T_c}
	T_c = \frac{\hbar|\omega_1|}{2\pi k_B}.
\end{equation}

Equation~(\ref{eq: T_c}) has the same form as the crossover temperature in particle systems~\cite{Benderskii_1994},
but there is a significant difference.
In the case of particle rearrangements, $\omega_1 = \sqrt{-\lambda_1}$
where $\lambda_1$ is the negative eigenvalue of the Hessian matrix at the first-order saddle point on the energy surface, 
whereas for magnetic systems, governed by the Landau-Lifshitz equation of motion, $\omega_1$ depends on the 
determinant of the Hessian matrix, i.e.~all the eigenvalues.
For a single spin described by spherical polar coordinates $\theta$ and $\phi$~\cite{Vlasov_2016}
\begin{equation}
\label{eq: 1-spin Tc}
	T_c = \frac{\sqrt{E^\dagger_{\theta\theta}E^\dagger_{\phi\phi} - \left(E^\dagger_{\theta\phi}\right)^2}}{2\pi k_B\mu\sin\theta^\dagger}\,,
\end{equation}
where $E^\dagger_{\theta\theta}$, $E^\dagger_{\phi\phi}$, and $E^\dagger_{\theta\phi}$ are second derivatives of the energy function. 
For a multi-spin system, it is better to work with Cartesian coordinates, as has been done in the present case, because of the problems that can arise if any of the spins points in the vicinity of the two poles. 


Once the onset temperature for tunneling has been found, the transition rate at that temperature can be estimated using HTST~\cite{Bessarab_2012}. 
The quantum mechanical effects on the transition rate can, however, be large already at T$_c$. For example, in atomic rearrangements involving 
chemical reactions or diffusion events, the transition rate has in some cases been found to be a couple of orders of magnitude larger than the classical rate
at T$_c$ \cite{Asgeirsson_2018}. Significant quantum effects can, therefore, be evident well above T$_c$.  



\section{Model}

The calculations presented here are based on an extended Heisenberg Hamiltonian for the system
\begin{equation}
	\mathcal{H}(\vec{\mathbf{s}})
	=
 		\sum\limits_{<{ij}>}\, 
	\left[ 
         \mathbf{D}_{ij} \cdot (\mathbf{s}_i\times\mathbf{s}_j)
         -
         J ~\mathbf{s}_i\cdot\mathbf{s}_j 
         \right]
	 - \sum\limits_{i=1}^{N} 
	 \left[ 
	 \mu ~{\bf B} \cdot \mathbf{s}_i + K  {s}_{i,z}^2
	 \right]
	\label{eq: Hamiltonian}
\end{equation}
where 
$\mathbf{D}_{ij}$ is the Dzyaloshinskii-Moriya vector lying in the plane of the lattice parallel to the vector pointing between two nearest neighbor sites $i$ and $j$,
thereby supporting Bloch type skyrmions,
$J$ is the exchange coupling parameter,
$K$ the out of plane anisotropy constant, and
$\bf B$ the uniform external magnetic field
 applied perpendicular to the lattice plane.
The sum 
$<{ij}>$
includes distinct nearest neighbor pairs. 
The system consists of 2500 spins on a triangular lattice with lattice spacing $\alpha=1$ and periodic boundary conditions. 
This model can, in particular, represent well the PdFe/Ir(111) system.
Parameter values obtained from density functional theory calculations~\cite{Malottki_2018} have been used in calculations of skyrmion lifetime using HTST 
and found to give results that are consistent with experimental measurements of PdFe/Ir(111)~\cite{Bessarab_2018}.


\section{Results and Discussion}

\subsection{Scan over parameter values}

\begin{figure}[t]
\centering    
		\includegraphics[width=0.65\linewidth]{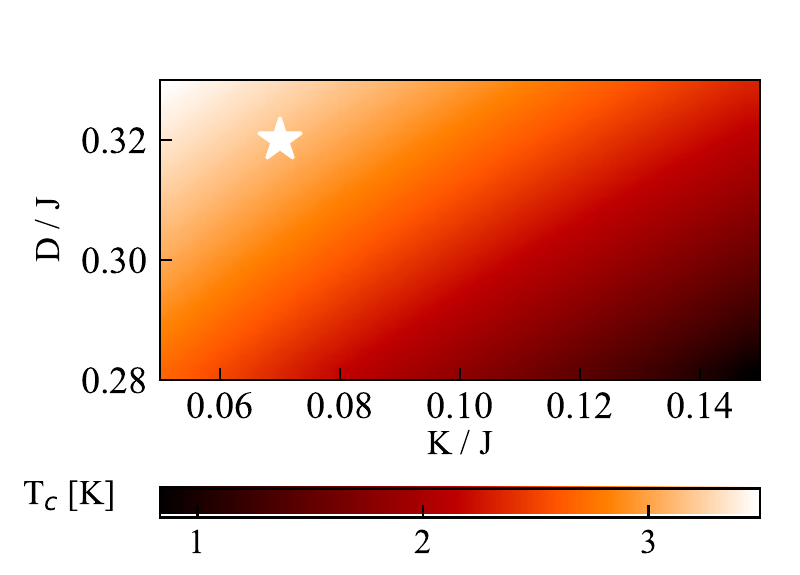}
\caption{
Onset temperature for quantum mechanical tunneling of a skyrmion in a two-dimensional triangular lattice
as a function of scaled parameters in the extended Heisenberg Hamiltonian: the Dzyaloshinskii-Moriya parameter $D$
and the anisotropy parameter $K$ in units of the exchange coupling parameter $J$. The scaled external magnetic field is
$B=0.73~B_D$, where $B_D= D^2/(\mu J)$ is the critical field.
The white star indicates the calculated value for the PdFe/Ir(111) system where $J$~=~7~meV~\cite{Romming_2013} and the field then corresponds to $B$~=~3~$T$, 
giving a crossover temperature of T$_c$~=~4~K. The lifetime of the skyrmion is, however, very long under those conditions as illustrated 
in Fig. 2, but can be reduced by increasing the strength of the external field (see Fig. 3).
}
	\label{fig1}
\end{figure}

Fig.~\ref{fig1} shows the calculated onset temperature, T$_c$, for the tunneling of the skyrmion to the ferromagnetic state
as a function of scaled Dzyaloshinskii-Moriya parameter, $D/J$, 
and scaled anisotropy parameter, $K/J$, when the applied magnetic field is $B=0.73 B_D$ where $B_D= D^2/(\mu J)$ is the critical field~\cite{Bogdanov_1994}. 
The value obtained for T$_c$ varies from 1 K to 4 K over the range of parameter values used.

\begin{figure}[t]
\centering
    \includegraphics[width=0.65\linewidth]{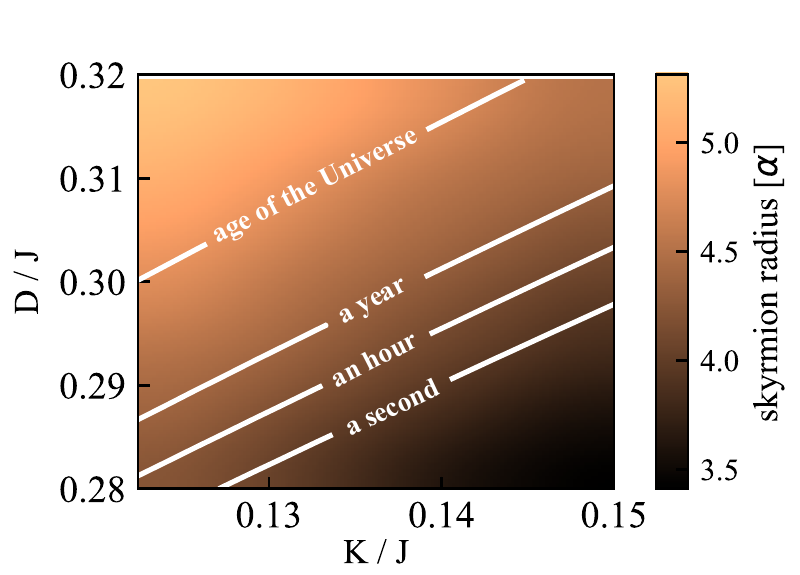}
    \caption{  
Calculated skyrmion lifetime at the onset temperature for quantum mechanical tunneling as a function of 
scaled parameters in the extended Heisenberg Hamiltonian: the Dzyaloshinskii-Moriya parameter $D$
and the anisotropy parameter $K$ in units of the exchange coupling parameter $J$
(white contour lines), 
superimposed on a contour graph of the 
skyrmion radius defined to be the distance from the skyrmion center to spins with $s_z$=0 
(see color bar to the right).
The calculations are for scaled external magnetic field $B~=~0.73~B_D$
and lifetime is evaluated for $J$~=~7~meV $B$~=~3~$T$~\cite{Romming_2013}.
For small skyrmions, the lifetime at the crossover temperature for tunneling, T$_c$, can be on laboratory scale. 
}
    \label{fig2}
\end{figure}

The shift in annihilation mechanism from over-the-barrier to tunneling could be observed by measuring the lifetime of skyrmions 
as a function of temperature. It is manifested by a break between a temperature dependent lifetime above T$_c$ to temperature independent lifetime below T$_c$.
The Arrhenius graph displaying the logarithm of the lifetime vs. inverse temperature shows an intersection between a straight line with slope determined by the 
activation energy above above T$_c$ to a line with zero slope below T$_c$.
In order for the crossover to be observable, the skyrmion lifetime at T$_c$ needs to be on laboratory time scale, i.e. on the order of seconds or minutes.
It is, therefore, important to also estimate the lifetime of the skyrmion at T$_c$ and this is done here using HTST.

Fig.~\ref{fig2} shows the calculated lifetime as a function of scaled parameter values in the Hamiltonian.
The size of the skyrmion is also indicated by the color scale.
The stability of skyrmions is to large extent related to their size~\cite{Varentsova_2018}, 
as can be seen from Fig.~\ref{fig2}. The smaller the skyrmion, the shorter the lifetime.
The lifetime changes remarkably strongly over this small range of parameter values.
Since the lifetime is also a strong function of temperature, the window for observing the crossover is limited to a narrow range 
of parameter values. But, this range does exist and it should be possible to find materials where such measurements can be performed.

\begin{figure}[t]
\centering
      \includegraphics[width=0.65\linewidth]{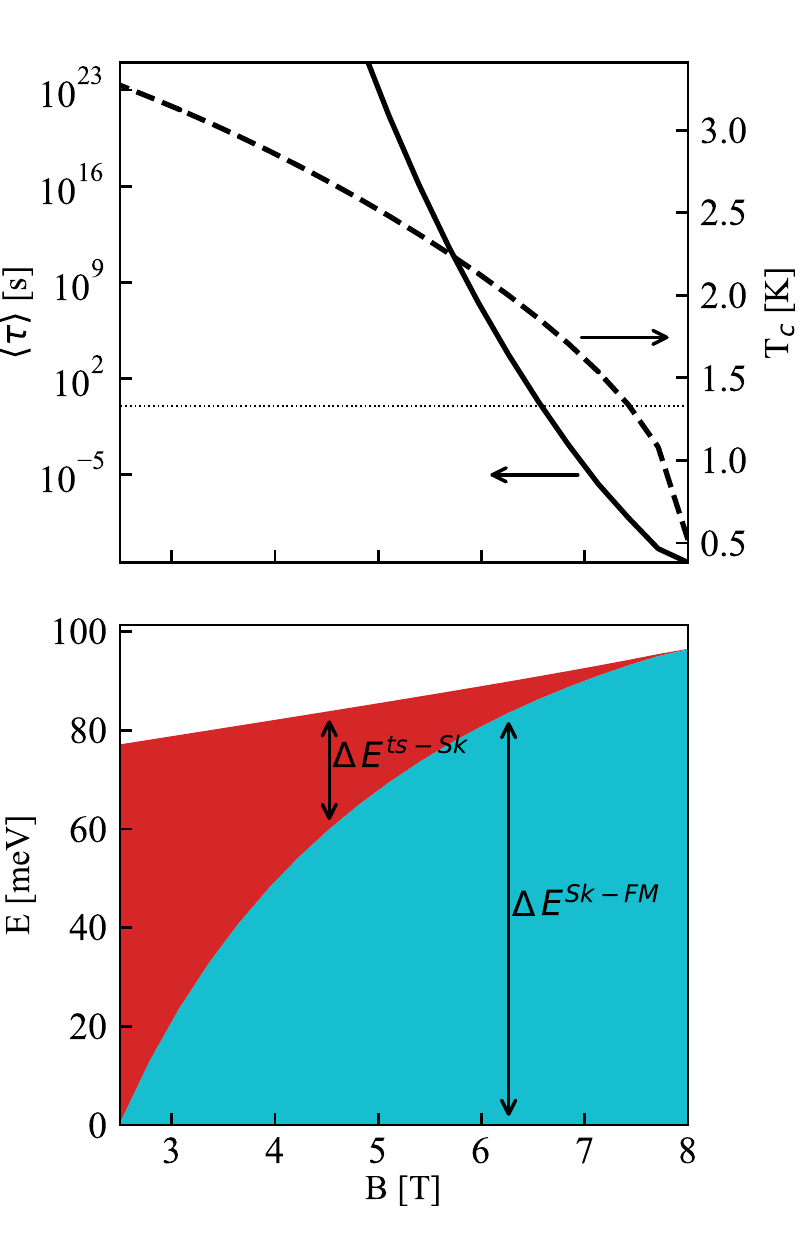}
    \caption{
Calculated results for a skyrmion in the PdFe/Ir(111) system as a function of applied magnetic field.
The parameters in the extended Heisenberg Hamiltonian are taken to be
$D/J$ = 0.32 and $K/J$ = 0.07
with
$J$ = 7 meV and $\mu$ = 3 $\mu_B$ (consistent with Ref.\,\cite{Romming_2013}, see also Fig. 1).
Upper: 
Onset temperature for tunneling, T$_c$, (dashed line, right axis) and
skyrmion lifetime in seconds at that temperature, $<\tau>$ (solid line, left axis).
In a field of B = 6.4~$T$, the lifetime of the skyrmion at the onset temperature for tunneling is predicted to be a couple of minutes,
and it drops to 10 seconds at 6.5~$T$.
The dotted line indicates a lifetime of 1 s.
Lower: 
Relative energy of the metastable skyrmion with respect to the ferromagnetic state,
$\Delta E^{Sk-FM}$, and energy barrier for the annihilation of the skyrmion, $\Delta E^{ts-Sk}$.
}
    \label{fig3}
\end{figure}

\subsection{The PdFe/Ir(111) system}

A great deal of experimental effort has focused on skyrmions in the PdFe/Ir(111) system~\cite{Romming_2013,Kubetzka_2017}.
Parameter values obtained from density functional theory~\cite{Malottki_2018} are found to give
results that are consistent with the experimental observations~\cite{Bessarab_2018}.  
Fig.~\ref{fig1} shows calculated results using those parameter values when the applied field is B = 3~$T$.
Under such conditions, the crossover temperature is 4 K but the lifetime is much too long at that temperature for experimental observations.
By increasing the applied magnetic field, the size of the skyrmion is reduced and the lifetime thereby shortened,
as illustrated in Fig.~3.
The variation of T$_c$ with respect to the strength of the anisotropy and the DMI is shown in more detail in Fig.~4. 
With an applied field of B = 6.4~$T$, the lifetime is brought down to a couple of minutes and the crossover temperature is still not too low, 
about 1.5 K. 
The lifetime is an extremely strong function of the field. In a field of 6.5~$T$ it is predicted to be 10 seconds. 
Our calculations therefore predict that tunneling of skyrmions could be observed in this system on a laboratory time scale by applying strong enough magnetic field. 
Other materials may of course be better suited for the observation of skyrmion tunneling, but the PdFe/Ir(111) system is at least one possible candidate.

The DMI and crystalline anisotropy are relatively strong as compared to the exchange interaction in PdFe/Ir(111), as shown in Fig. 5. 
Therefore, they cannot be considered as small perturbations for this system.
Also, even when the size of the skyrmion is reduced by applying a field of B = 8~$T$, 
the exchange interaction has not reached the BP limit of 4$\pi \sqrt{3} J$. 
The BP soliton is, therefore, not a good approximation for this system and the more general methodology presented here without pre-determined 
shape function is needed in order to obtain accurate results. 


%
\begin{figure}[t]
\centering
      \includegraphics[width=0.65\linewidth]{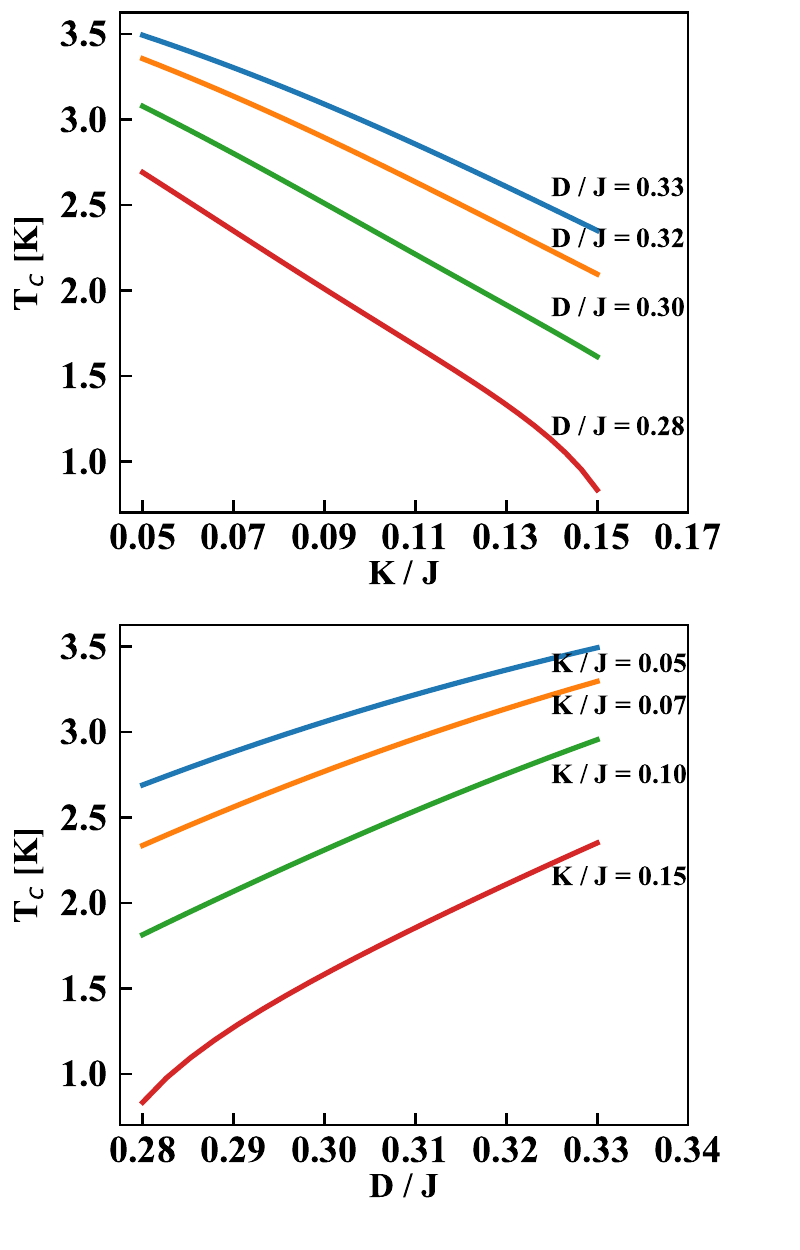}
    \caption{
Calculated onset temperature for tunneling of a skyrmion, T$_c$, in PdFe/Ir(111) as a function of anisotropy parameter, $K/J$, (upper)
and as a function of Dzyaloshinskii-Moriya parameter $D/J$ (lower), in both cases scaled by the exchange parameter, $J$.
}
    \label{fig4}
\end{figure}

\begin{figure}[t]
\centering
      \includegraphics[width=0.65\linewidth]{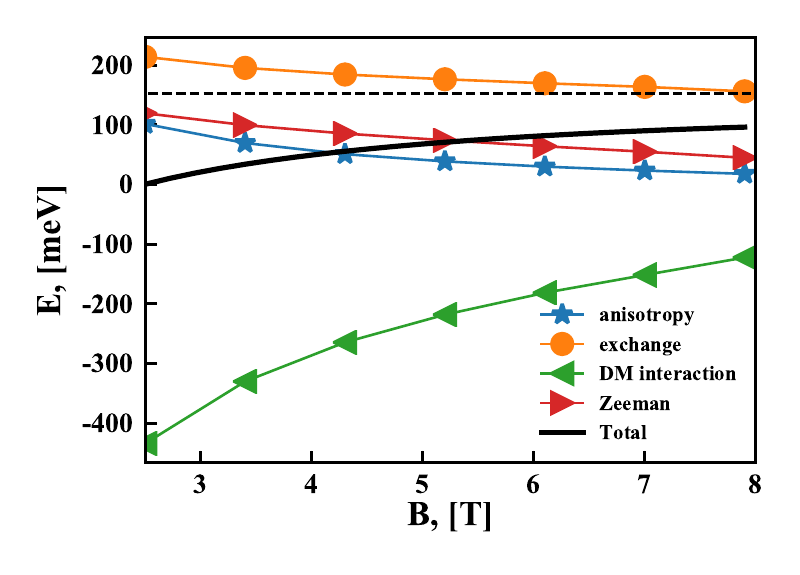}
    \caption{
Various contributions to the energy of a skyrmion in PdFe/Ir(111) 
with respect to the energy of the ferromagnetic state
as a function of applied magnetic field in units of Tesla.
Even for a field of B = 8~$T$, the DMI has a larger magnitude than the exchange interaction and the latter has not reached the
Belavin-Polyakov limit of 4$\sqrt{3}\pi J$ (indicated by a dashed line).
}
    \label{fig5}
\end{figure}


\section{Conclusions}

The calculations presented here make a prediction regarding conditions and materials parameters that make it possible to observe  
a crossover from classical over-the-barrier to quantum mechanical tunneling mechanism of skyrmion collapse,
an interesting example of macroscopic tunneling. It is important to assess to what extent 
the tunneling mechanism needs to be taken into account when estimating the lifetime of skyrmions for practical applications.
The method presented here includes all spins in the system without assuming a preconceived effective shape of the skyrmion
or a predefined collective reaction coordinate. The onset temperature is estimated using an instanton approach and the lifetime
of the skyrmion at the crossover temperature estimated from the harmonic approximation of transition state theory. 
This approach has previously been shown to work well in calculations of tunneling of a macrospin representing a molecular magnet
but is extended here to systems of multiple spins.


\vskip 0.6 true cm

\section*{Acknowledgments}

This work was supported by the Icelandic Research Fund, 
the Research Fund of the University of Iceland, 
the Russian Foundation of Basic Research (grants RFBR 18-02-00267 and 19-32-90048), and 
the Foundation for the Advancement of Theoretical Physics and Mathematics ``BASIS'' under Grant No. 19-1-1-12-1.
PFB thanks the Alexander von Humboldt Foundation for support.


\vskip 1 true cm

\vfill\eject

\end{document}